\documentclass[pra,showpacs,twocolumn,byrevtex,superscriptaddress]{revtex4}
\usepackage{graphicx}
\usepackage{subfigure}
\usepackage{dcolumn}
\usepackage{amsmath}
\usepackage{epsfig}
\usepackage{color}
\usepackage{amsmath}

\usepackage{multirow}

\begin{document}
\title{Secure quantum communication in the presence of phase- and polarization-dependent loss}

\author{Chenyang Li}
\email{chenyangli@ece.utoronto.ca}
\affiliation{Center for Quantum Information and Quantum Control, Department of Electrical \& Computer Engineering
and Department of Physics, University of Toronto, Toronto, Ontario, M5S
3G4, Canada}

\author{Marcos Curty}
\affiliation{Escuela de Ingenier\'{i}a de Telecomunicaci\'{o}n, Department of Signal Theory and Communications, University of Vigo, Vigo E-36310, Spain}

\author{Feihu Xu}
\affiliation{Shanghai Branch, Hefei National Laboratory for Physical Sciences at the Microscale, University of Science and Technology of China, Shanghai 201315, China}

\author{Olinka Bedroya}
\affiliation{Center for Quantum Information and Quantum Control, Department of Electrical \& Computer Engineering
and Department of Physics, University of Toronto, Toronto, Ontario, M5S 3G4, Canada}
\author{Hoi-Kwong Lo}
\email{hklo@ece.utoronto.ca}
\affiliation{Center for Quantum Information and Quantum Control, Department of Electrical \& Computer Engineering
and Department of Physics, University of Toronto, Toronto, Ontario, M5S 3G4, Canada}

\date{\today}
%\maketitle

\begin{abstract}

Silicon photonics holds the promise of the miniaturization of quantum communication devices. Recently, silicon chip optical transmitters for quantum key distribution (QKD) have been built and demonstrated experimentally. Nonetheless, these silicon chips suffer substantial phase- and polarization-dependent loss (PDL) which, if
unchecked, could compromise the security of QKD systems because of overestimating the secret key rate. Here, we first  restore the security by regarding the single photons without phase and polarization dependence as untagged and secure qubits. Next, by using a post-selection technique,
one could implement a secure QKD protocol that provides a high key generation rate even in the
presence of  severe phase and polarization dependent loss.
Our solution is simple to realize in a practical experiment as it does not require any hardware modification.

%\vspace{0.8cm}
\end{abstract}

\maketitle

\section{introduction}\label{Sec1}
Quantum key distribution (QKD) allows two distant parties, Alice and Bob, to share a common string of secret data \cite{Lo2014}. Based on the laws of quantum mechanics, QKD offers unconditional security. Recently, secure QKD has been experimentally demonstrated over 404 km \cite{Yin2016} of telecom fiber, based on the measurement-device-independent quantum key distribution (MDI-QKD) protocol \cite{Lo2016}. A quantum communication satellite has been launched by China demonstrating satellite-based distribution of entangled photon pairs over 1200 km \cite{Yin2017}. In the long term, quantum communication promises to offer security for both civilian and government applications. It is widely believed that CMOS-compatible silicon photonics holds the potential to  dramatically lower the cost of QKD devices, thus bringing QKD to widespread applications. Indeed, a seminal proof-of-principle QKD experiment with a silicon photonic chip transmitter has been recently implemented in \cite{Ma2016}. More recently, a Bristol group \cite{Sibson2017} has performed another QKD experiment also with a silicon photonic chip transmitter. These two experiments highlight the potential of silicon photonics in quantum communication.

The most common method of achieving fast modulation in silicon devices so far is to exploit the plasma dispersion effect \cite{Reed2010}, in which the concentration of free charges in silicon changes
both the real and imaginary parts of the refractive index of the material, which affect the phase and intensity of the propagating light simultaneously. On the other hand,
an important assumption in many security proofs of QKD is that the intensity of a quantum signal is independent of the actual quantum state encoded \cite{Hwang2003,wang2005,Lo2005,Ma2005}. This is to avoid the existence of side channel information that could allow Eve to learn the photon polarization through an intensity measurement. Unfortunately, this important assumption in the security proofs of QKD {\it cannot} be guaranteed by current silicon photonics modulators. For instance, in the Bristol group's silicon chip QKD transmitter \cite{Sibson2017}, the magnitude of the phase dependent loss was apparently measured to be about 1 dB whereas the polarization dependent loss was about 1.6 dB in the silicon chip transmitter demonstrated in \cite{Ma2016}.
Polarization dependent loss is also an important issue when four separate lasers are used to implement the BB84 protocol. In satellite-based QKD,   eight laser diodes are integrated inside a single transmitter---four for
signal and four for decoy states, emitting photons in a preset polarization state \cite{Liao2017,Bedington2017}. A possible drawback of this approach is that the intensities of the light pulses emitted from the diodes are not always identical.
Such  polarization dependent loss,
if unchecked, could reduce the secrete key rate and furthermore render a QKD protocol insecure. In other words, if Alice and Bob are unaware of the existence of the PDL, by overestimating their key rate, they will generate a key that is too long and is, therefore, not guaranteed to provide information-theoretic security.
The main goal of this paper is to restore the security
of QKD in the presence of phase or polarization dependent loss.

Generally speaking, our view is that there are two potential methods
to restore the security of QKD in the presence of PDL.
The first method is hardware based. The sender, Alice, may add
another intensity modulator to compensate for PDL and ensure that
the signal intensity is independent of the actual polarization/phase state.
Such a method is theoretically simple, but may add
further complexity to an experiment by requiring an additional
component, which may in itself introduce new imperfections.
In this paper, we consider a second method, which is software based.
Our idea is to modify the security proof and develop a software-based solution
to compensate for PDL. In particular, we will show how, through
post-selection of signals, we are able to restore the security of
decoy-state based QKD and maximize its secret key generation rate.

In the bigger context, our work serves as an example to demonstrate the power of
software solutions in QKD system designs and security analyses in the
presence of device imperfections.

This paper is organized as follows.
In Sec.\ref{Sec2}, we explain the
physical origin of PDL in silicon chip
QKD transmitters.
In Sec.\ref{Sec3}, we present a theoretical model for the security analysis
of decoy-state QKD with PDL.
In Sec.\ref{Sec4}, we present a security proof for the QKD system with PDL and we propose a post-selection scheme to maximize the key generation rate.
In Sec.\ref{Sec5}, we perform a numerical optimization to maximize the secret key rate  by considering the asymptotic  case where Alice sends Bob an infinite number of signals.
In Sec.\ref{Sec6}, we introduce a technique to apply the refined data analysis to the QKD system.
In Sec.\ref{Sec7}, we analyze the finite key reginme
in QKD and show that our results are robust in the practical setting of
an experiment with a finite data size.  Finally, in Sec.\ref{Sec8}, we provide some
concluding remarks.

\section{Phase/polarization dependent loss}
\label{Sec2}

Let us first review the physical origin of PDL in a silicon photonics transmitter for QKD. The plasma dispersion effect is widely used to achieve fast modulation in silicon devices. Injection or depletion of free carriers in silicon changes the real and imaginary parts of the refractive index of the material, which then change the phase and absorption of the propagating light. Soref and Bennett \cite{soref1987} experimentally quantified the  refractive index change over a wide range of electron and hole densities. For instance, at a wavelength of $1.55 \mu m$ the changes in the real part $\bigtriangleup n$ of the refractive index and in the absorption coefficient $ \bigtriangleup \alpha$ over the carrier densities in silicon can be expressed as:
\begin{gather}\label{eq1}
  \bigtriangleup n=-[8.8\times10^{-22}\times \bigtriangleup N_e+8.5\times10^{-18}\times \bigtriangleup N_h^{0.8}],  \notag \\
  \bigtriangleup \alpha=8.5\times10^{-18}\times \bigtriangleup N_e+6.0\times 10^{-18}\times \bigtriangleup N_h ,
\end{gather}
where $\bigtriangleup N_e$ and $\bigtriangleup N_h$ are, respectively, the changes in the free electron density and free hole density.

As already mentioned above, the overall PDL in the QKD silicon photonic transmitter implemented in \cite{Ma2016} was found to be around 1.6 dB. Similarly, the Bristol group  \cite{Sibson2017}  found that PDL provides stringent restrictions for high speed modulators, like carrier-injection or carrier-depletion modulators. In their experiment, PDL was
found to be about 1 dB. In summary, PDL was non-negligible in both experimental demonstrations.
 Note that in \cite{Ma2016,Sibson2017}, the authors considered PDL in the silicon chip only, and ignored PDL in other fiber components. This is reasonable because, for instance, according to \cite{Pdltest},  the typical PDL in a single-mode fiber channel at 10 km is less than 0.05 dB, which is
negligible.  We have also performed a simple measurement of PDL in a fiber-based polarization encoding QKD source. The schematic set-up is shown in Fig.1. The output power of the system is measured for different polarization states, which are created by sweeping the voltage applied to the polarization modulator. As shown in Fig.1, the ratio between the maximum and minimum power was found to be around 0.2 dB, which is rather small.
  In the next section, we  discuss  how the existence of PDL in
 the state preparation process of QKD could affect its security.

\begin{figure}[!htb]
\centering
\subfigure[]{\label{a}
\includegraphics [width=77mm,height=44mm]{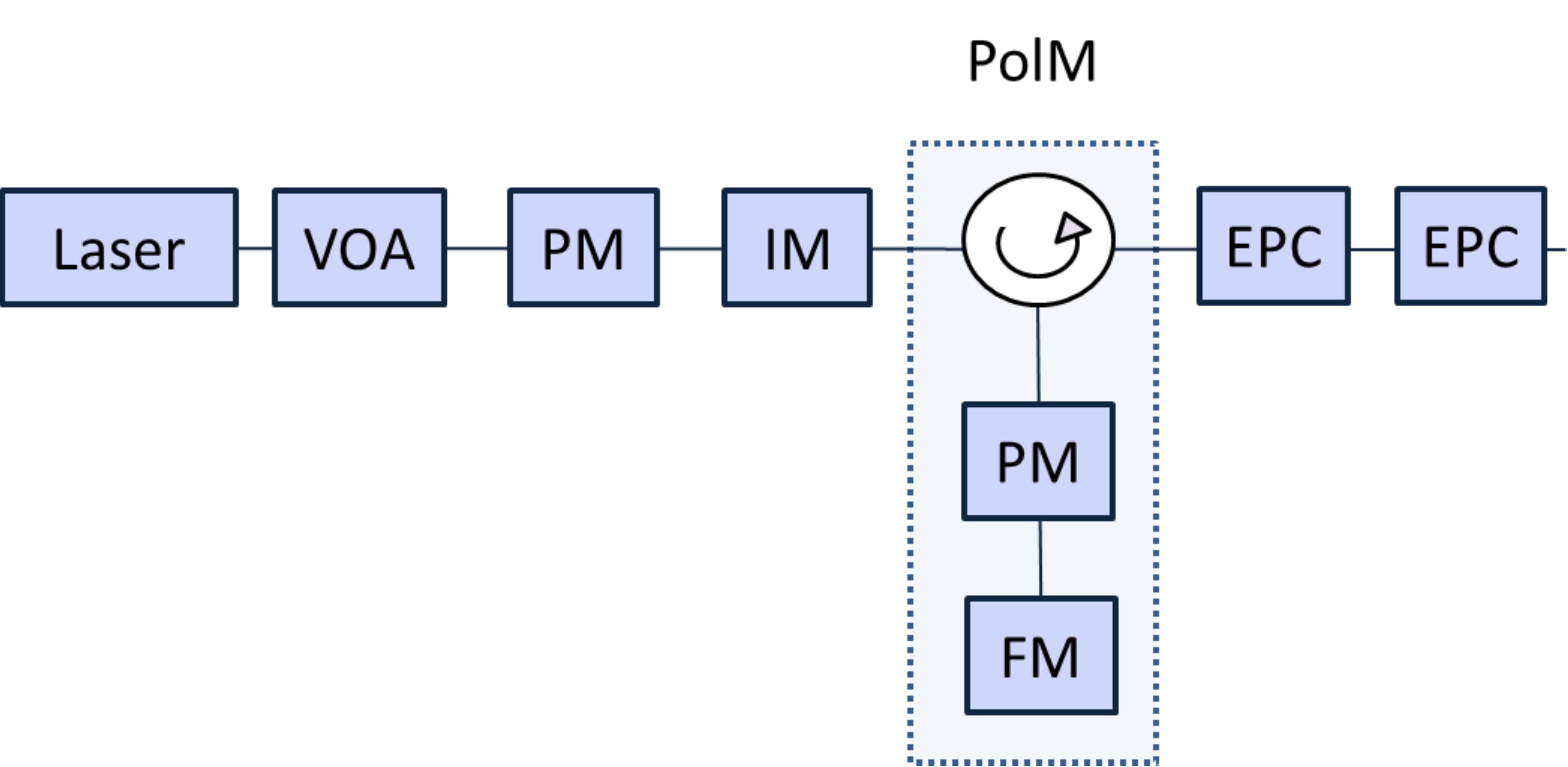}}
\subfigure[]{\label{b}
\includegraphics [width=80mm,height=55mm]{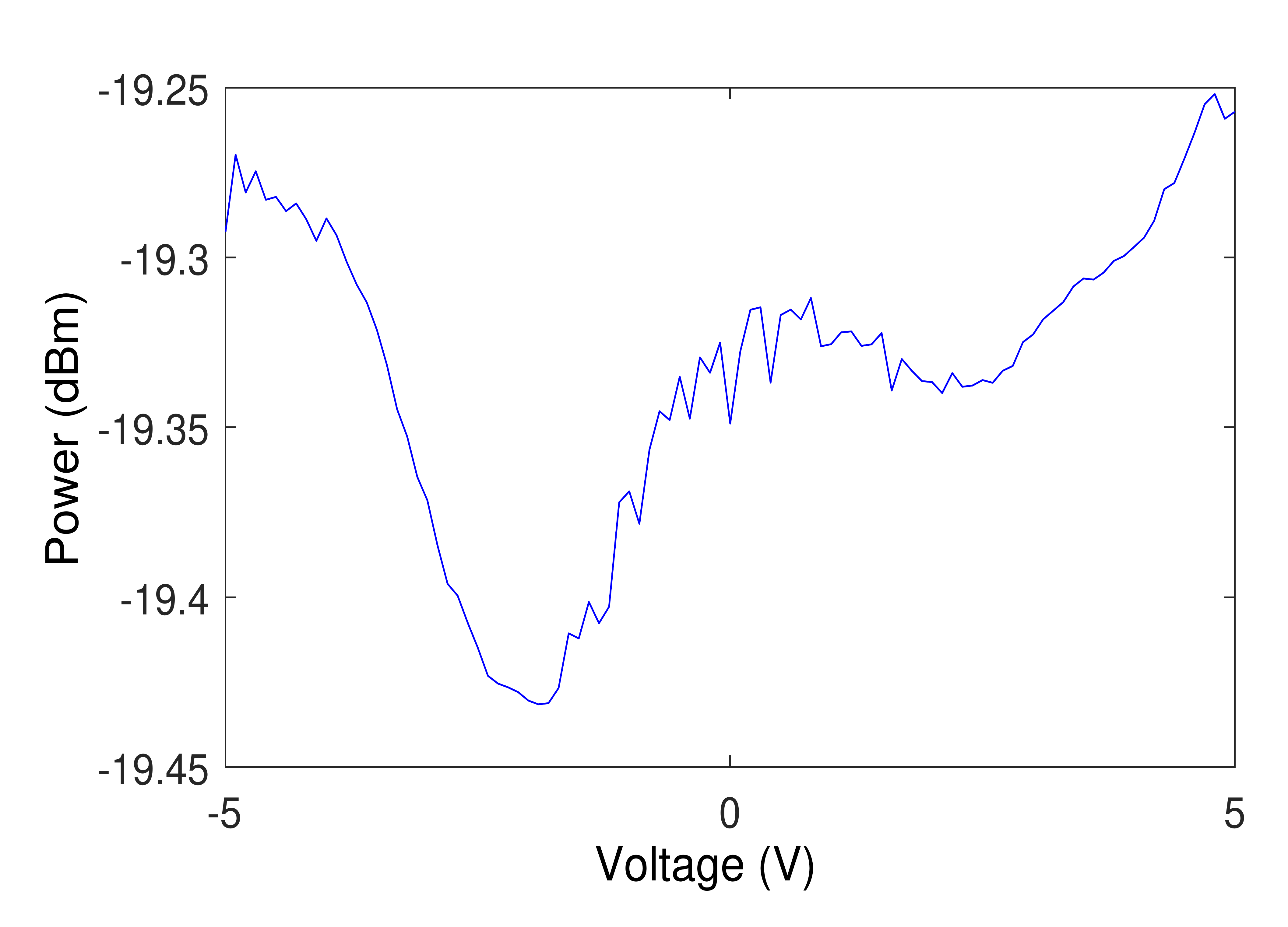}}

\caption{(a) Schematic of the QKD source used for the PDL measurement. A CW laser  is attenuated using a variable optical attenuator (VOA).  The setup further consists of a phase modulator (PM) for phase randomization and an intensity modulator (IM) to generate pulses. The polarization of the weak coherent pulses is modulated by a polarization modulator (PolM) which is based on a structure proposed in \cite{Martinez2009} using a Faraday Mirror (FM). Two electronic polarization controllers (EPCs) are included in the source for polarization alignment. (b) The variation of power at the output of the source with respect to the voltage applied to the phase modulator in the polarization modulator. The maximum variation of the output power for different polarization states is around $0.2$ dB. }
\end{figure}\label{Fig1}

\section{Model}
\label{Sec3}

To better understand how PDL affects the security of QKD, we  first present a mathematical model that describes the typical behaviour of the different elements of a QKD system. In particular, we  model the source, channel, detector, yield and  quantum bit error rate in
decoy-state QKD. After that, we explain how PDL affects the security analysis of QKD.

 \subsection{State preparation}\label{Sec3.1}
Here, a strongly attenuated laser pulse is modeled as a weak coherent state.  For concreteness, below we shall focus on polarization encoding, but similar arguments can be applied to phase encoding as well.
Since the intensity of a pulse may now depend on its polarization, we denote  the expected photon numbers of the signal states for horizontal, vertical, diagonal (45-degrees), and antidiagonal (135-degrees) polarizations by $\mu_{s,H},\  \mu_{s,V},\ \mu_{s,D}$   and $ \mu_{s,A}$ respectively.
Assuming that the phase of each pulse is totally randomized, the density matrix which describes the state emitted by Alice is given by \cite{Ma2005,Curty2010}:

\begin{equation}\label{eq2}
  \rho_{s,M} =\sum_i \frac{\mu_{s,M}^i}{i!}e^{-\mu_{s,M}}|i\rangle\langle i|_M,
\end{equation}
where $M$ represents one of the four polarizations with $M \in \{H,V,D,A\}$.
Note that such a phase randomized weak coherent pulse is a mixture of Fock states whose photon number
is Poissonian distributed with mean $\mu_{s,M}$.

For simplicity and for the moment we shall assume  that PDL only occurs in the horizontal and diagonal polarizations. This means that:
\begin{equation}\label{eq3}
    \mu_{s,V}= \mu_{s,A}= \mu_{s},
\end{equation}
where $\mu_{s}$ is the original intensity selected by Alice.
Also, for simplicity, we shall consider the special case where the same PDL occurs in the two bases:
\begin{gather}
  \mu_{s,H}=L\times \mu_{s,V}, \notag \\
  \mu_{s,D}= L\times \mu_{s,A},\label{eq4}
\end{gather}
where $L$ is  the loss coefficient and has the form $L=10^{-PDL/10}$ with the parameter $PDL$ measured in dB.

Note, however, that the method that we present later on to achieve secure QKD in the presence of PDL is rather general
and does not rely on the  assumptions given by Eqs. (3)-(4).

\subsection{Transmission and detection efficiency}\label{Sec3.2}

For a fiber-based channel, the transmittance $\eta_{channel}$ can be expressed as \cite{Ma2005,Curty2010}:
\begin{equation}\label{eq5}
  \eta_{channel}=10^{-\alpha d/10},
\end{equation}
where $\alpha$ represents the loss coefficient of the channel measured in dB/km, and $d$ is the transmission distance measured in km.
The overall transmission and detection efficiency between Alice and Bob, $\eta_{sys}$, can be written as:
\begin{equation}\label{eq6}
  \eta_{sys}=\eta_{channel}\eta_{Bob} ,
\end{equation}
where  $\eta_{Bob}$ denotes the overall transmittance of Bob's detection apparatus, which includes the detection efficiency of his detectors.

\subsection{Yield and quantum bit error rate}\label{Sec3.3}

Since now the intensity of the signal state depends on the polarization,
 we apply a refined data analysis scheme to analyze the parameters for different polarizations.
Thus, we define $Y_i$ to be the yield of an $i$-photon state (i.e., the probability to observe a detection click at Bob's side), and $ Y_{i,M}$ with $M \in \{H,V,D,A\}$ to be the yield of an $i$-photon state prepared in the given polarization.
Note that in a single mode fiber without PDL, we assume:
\begin{equation}\label{eq7}
  Y_i=Y_{i,H}=Y_{i,V}=Y_{i,D}=Y_{i,A}.
\end{equation}
The yield of an $i$-photon state originates from two main contributions: the background rate $Y_0$ and the signal states. For a typical channel model, the yields $Y_i$ can be expressed as \cite{Ma2005, Curty2010}.
\begin{equation}\label{eq8}
   Y_i=1-(1-Y_0){(1-\eta_{sys})}^i.
\end{equation}
The gains for different polarizations (i.e., the overall probabilities to observe a detection click at Bob's side) are given by:
\begin{equation}\label{eq9}
 Q_{s,M}=\sum_i \frac{\mu_{s,M}^i}{i!}e^{-\mu_{s,M}}Y_{i,M}=1-(1-Y_{0,M})e^{-\mu_{s,M}\eta_{sys}},
\end{equation}
where $M \in \{H,V,D,A\}$. Note that the value of the different gains is directly experimentally observed in a QKD run.

Then, we define $e_i$ to be the error rate of an $i$-photon state, and $ e_{i,M}$ with $M \in \{H,V,D,A\}$ to be the error rate of an $i$-photon state  prepared in the given polarization. For simplicity, in a single mode fiber without PDL we also assume:
\begin{equation}\label{eq10}
  e_i=e_{i,H}=e_{i,V}=e_{i,D}=e_{i,A}.
\end{equation}
For a typical channel model, the $i$-photon error rate $e_i$ is  given by \cite{Lo2005, Ma2005}:
\begin{equation}\label{eq11}
  e_i=\frac {Y_0e_0+(Y_i-Y_0)e_d}{Y_i},
\end{equation}
where $e_d$ is the probability that a signal hits an erroneous detector due to
the misalignment in the
quantum channel. For simplicity, we assume that $ e_d $ is independent of the distance and $e_0=1/2$ (i.e., we consider that the background is random).

The overall quantum bit error rates (QBERs), $E_{s,M}$ with $M \in \{H,V,D,A\}$,  for different polarizations are given by:
\begin{multline}\label{eq12}
 E_{s,M}=\frac{1}{Q_{s,M}}\sum_i \frac{\mu_{s,M}^i}{i!}e^{-\mu_{s,M}}Y_{i,M}e_{i,M} \\
 =\frac{1}{Q_{s,M}}(Y_{0,M}(e_{0,M}-e_d)+e_d[1-(1-Y_{0,M})e^{-\mu_{s,M,}\eta_{sys}}]),
\end{multline}
The value of the different QBERs given by Eq.(\ref{eq12}) is also directly experimentally observed in a QKD run.

Note that the assumption about polarization independence of the quantum channel,
as described in Eqs.(\ref{eq7}) and (\ref{eq10}), is used to simplify our numerical
simulations and discussions.
However,  the security analysis that we present later on is general and does not require such an assumption.

\section{ Security proof and Post-selection scheme }\label{Sec4}
\subsection{General security proof for QKD with PDL}
In this section, we  restore the security of decoy-state QKD in the presence of PDL at the source.
But before we do so, let us first consider the secret key rate of the decoy-state BB84 protocol without PDL.
Based on the security analysis presented in \cite{Lo2005, Ma2005}, which combines the idea of the entanglement distillation approach by Gottesman-Lo-L\"{u}tkenhaus-Preskill in \cite{Gottesman2004} with the decoy state method, the secret key rate formula, in the asymptotic limit of
infinitely many quantum transmission data, can be written as:
\begin{equation}\label{eq13}
  R \geq q\{Q_1[1-H(e_{1,phase})]-Q_sf(E_s)H(E_s)\},
\end{equation}
where $q$ is the efficiency of the procotol ($q=1/2$ for the standard BB84 protocol \cite{BB84}, and $ q\approx 1$ for its efficient version \cite{Lo2005b}), $Q_s$ is the gain of the signal states, $E_s$ is the overall quantum bit error rate of the signal states, $Q_1$ is the gain of the
single-photon states, $e_{1,phase}$ is the phase error rate of the single-photon states, $f(E_s)$ is the efficiency of the error correction protocol, and $H(x)=-x{\rm log}_2(x)-(1-x){\rm log}_2(1-x)$ is the binary Shannon entropy function.

In the presence of PDL, now the intensity of the signal states depends on their actual polarization. That is, given a signal
state, the probability that it is a single photon state now depends on its
polarization. This violates a
fundamental assumption in the security proof of QKD---that the density matrices of the single-photon
components should be maximally mixed. i.e., the single-photon components
should be equally likely to be in the state  associated to a bit value zero and to a bit value one.
For this reason, the secret key rate stated in Eq.(\ref{eq13}) is not valid.

One simple  way to recover a valid secret key
rate in the presence of PDL is to replace $Q_1$ by
the gain of those single photons with a random choice of polarization. In other words, the secret key is only generated from the single-photon components whose density matrices are maximally mixed.
Then one obtains:
\begin{gather}\label{eq1103}
   Q_1 \longrightarrow   \min\{\mu_{s,H}e^{-\mu_{s,H}},\ \mu_{s,V}e^{-\mu_{s,V}}\} \times  Y_1,\\ \notag
   Y_1 \longrightarrow  \frac{Y_{1,H}+Y_{1,V}}{2}, \\ \notag
   e_{1,phase} \longrightarrow \frac{Y_{1,D}e_{1,D}+Y_{1,A}e_{1,A}}{Y_{1,D}+Y_{1,A}}, \\ \notag
   Q_s \longrightarrow \frac{1}{2}Q_{s,H}+\frac{1}{2}Q_{s,V}, \\ \notag
   Q_sE_s \longrightarrow \frac{1}{2}Q_{s,H}E_{s,H}+\frac{1}{2}Q_{s,V}E_{s,V}.
\end{gather}
Note that the gains and QBERs, $Q_{s,H},\ Q_{s,V},\ E_{s,H},\ E_{s,V}$, can  be experimentally obtained from a QKD run.
The phase error, $e_{1,phase}$, of the single-photon  components whose density matrices are maximally mixed  can be estimated as an average QBER of the single-photon states in the $X$ basis (diagonal and antidiagonal polarizations). Also, the yield of the single-photon components, $Y_1$, the gain of the signal, $Q_s$, and the QBERs, $E_s$, can be estimated as an average in the $Z$ basis (horizontal and vertical polarizations).

The main idea of this security proof is that,  if there are more vertically polarized single photons than horizontally polarized
single photons in the source, those excess photons are simply discarded and make no contribution to the
secret key. Besides, we still need to correct the errors caused by those excess signals. This method works well  if the PDL is small. When PDL is too large, the error corrections is very inefficient, so we introduce the post-selection scheme to solve this problem in the next section.

\subsection{Post-selection scheme for QKD with large PDL}
To obtain a higher secret key rate in the presence of large PDL, we now introduce a post-selection scheme.
As already mentioned above (see Eq.(4)), for simplicity let us consider that  the PDL occurs in the horizontal polarization. In this scenario, we want to keep the data from the horizontal  polarization because they
are fewer and thus precious to us. For the vertical polarization, we randomly keep or discard a portion of the signals to balance out its number with that of the horizontally polarized signals. For instance,
this can be done by introducing a post-selection probability $P$ such that
for each vertically polarized signal sent by Alice, she keeps it with  probability $P$ and discards it  with
probability $1-P$. Once Bob has measured all the incoming signals, Alice informs him which signals are kept and which ones are discarded.
By using such broadcast information, Alice and Bob can simply discard all the excess signals.
This means that they no longer need to correct the bit errors in these excess signals.
Therefore, they save the cost of error correction. This allows them to
obtain a higher secret key rate over  longer distances than what they would otherwise
have obtained. More concretely, with the post-selection scheme, we have that:
\begin{gather}\label{eq1103}
   \tilde{Q}_1 \longrightarrow   \min\{\mu_{s,H}e^{-\mu_{s,H}},\ P\times\mu_{s,V}e^{-\mu_{s,V}}\} \times  Y_1,\\ \notag
   Y_1 \longrightarrow  \frac{Y_{1,H}+Y_{1,V}}{2}, \\ \notag
   e_{1,phase} \longrightarrow \frac{Y_{1,D}e_{1,D}+Y_{1,A}e_{1,A}}{Y_{1,D}+Y_{1,A}}, \\ \notag
   \tilde{Q}_s \longrightarrow \frac{1}{2}Q_{s,H}+P\times\frac{1}{2}Q_{s,V}, \\ \notag
   \tilde{Q}_s\tilde{E}_s \longrightarrow \frac{1}{2}Q_{s,H}E_{s,H}+P\times\frac{1}{2}Q_{s,V}E_{s,V}.
\end{gather}
Note that the $e_{1,phase}$ and $Y_1$ are not modified, because one always requires that the density matrices of the single-photon components are maximally mixed. If $P=1$, it means that we do not apply post-selection  and keep all data.
With the post-selection scheme, the secret key rate can be expressed as:
\begin{equation}\label{eq14}
  R \geq q\{\tilde{Q}_1[1-H(e_{1,phase})]-\tilde{Q}_sf(\tilde{E}_s)H(\tilde{E}_s)\}.
\end{equation}

A key advantage of this approach is moreover that the post-selection can be done as a software solution.
In particular, there is no need to modify the hardware of the quantum transmission in
a QKD system. Since QKD hardware is typically more expensive than software and more difficult to modify, such a software solution is often preferred over a hardware solution.

\subsection{Optimal $P$ and $\mu_s$ }\label{Sec4.2}

The next question is: what is the optimal value of the post-selection probability $P$?
To answer this question,
here we discuss about how to optimize the secret key rate by choosing a post-selection probability.
Heuristically, since the goal of the post-selection scheme is to discard the excess signals,
one would expect that after
the post-selection, for the single-photon components, there is an equal probability that they are
vertically or horizontally polarized. This means that we theoretically expect that the value of $P$  satisfies the
following equations:
\begin{gather}
  \mu_{s,H}e^{-\mu_{s,H}} = P \times  \mu_{s,V}e^{-\mu_{s,V}}, \notag \\
  \mu_{s,H}=L  \mu_{s,V}\label{eq15},
\end{gather}
from which we obtain:
\begin{equation}\label{eq16}
  P_{optimal} = \frac {Le^{-L\mu_{s,V}}}{e^{\mu_{s,V}}}.
\end{equation}

\begin{figure}[htb]
\centering
 \includegraphics[width=75mm,height=50mm]{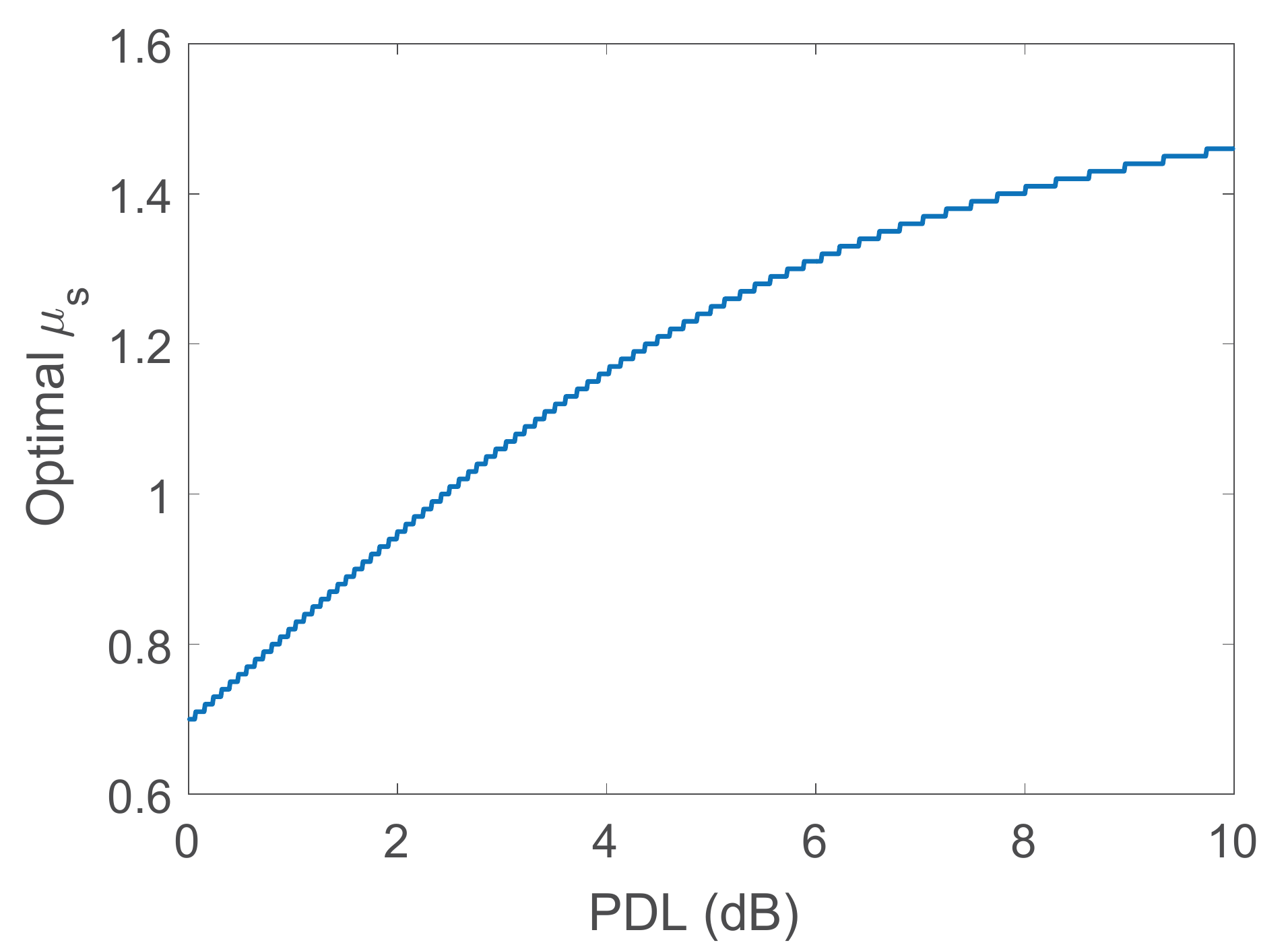}

\caption{By numerical simulation, we obtain the optimal value of $\mu_s$ that satisfies Eq.(\ref{eq18}). When PDL increases, the optimal $\mu_s$ increases as well in order to generate more single photons. The experimental parameters are given in Table \ref{table1}. }\label{Fig2}
\end{figure}\label{fig2}

If we consider the case where the background rate is low ($Y_0<< \eta_{sys}$) and the transmittance is small ($\eta_{sys}<<1$), then with the $P_{optimal}$  and the channel model described in the previous section, the key rate is given by \cite{Ma2005}:
\begin{gather}
   R\approx -\left(\frac{1}{2}P_{optimal}\eta_{sys}\mu_s+\frac{1}{2}\eta_{sys}L\mu_s\right)f(e_d)H(e_d) \notag \\ +\eta_{sys}L\mu_se^{-L\mu_s}(1-H(e_d)). \label{eq17}
\end{gather}
In this case, it can be shown that the rate is maximized if we choose a $\mu_s$ which satisfies:
\begin{multline}\label{eq18}
   [1-H_2(e_d)][Le^{-L\mu_s}-L^2\mu_se^{-L\mu_s}]=\\
   \frac{1}{2}f(e_d)H_2(e_d)[L+Le^{(1-L)\mu_s}+L\mu_se^{(1-L)\mu_s}(1-L)].
\end{multline}
By using the experimental parameters shown in Table \ref{table1}, we find that when PDL increases, $\mu_s$ increases as shown in Fig.\ref{Fig2}. We should note that when the value of PDL is larger than 4 dB, the optimal $\mu_s$ may exceed $1$. In the QKD scenario without PDL \cite{Lo2005, Ma2005}, the optimal $\mu_s$ always lies within the range of 0 and 1.
In the next section, we  numerically search the optimal values of  $P$ and $\mu_s$.
Our numerical simulation shows that our above heuristical argument is essentially
correct.

\section{infinite decoy states and infinite data sizes}\label{Sec5}
\begin{figure}[htb]
\centering
\subfigure[]{
 \includegraphics [width=80mm,height=48mm]{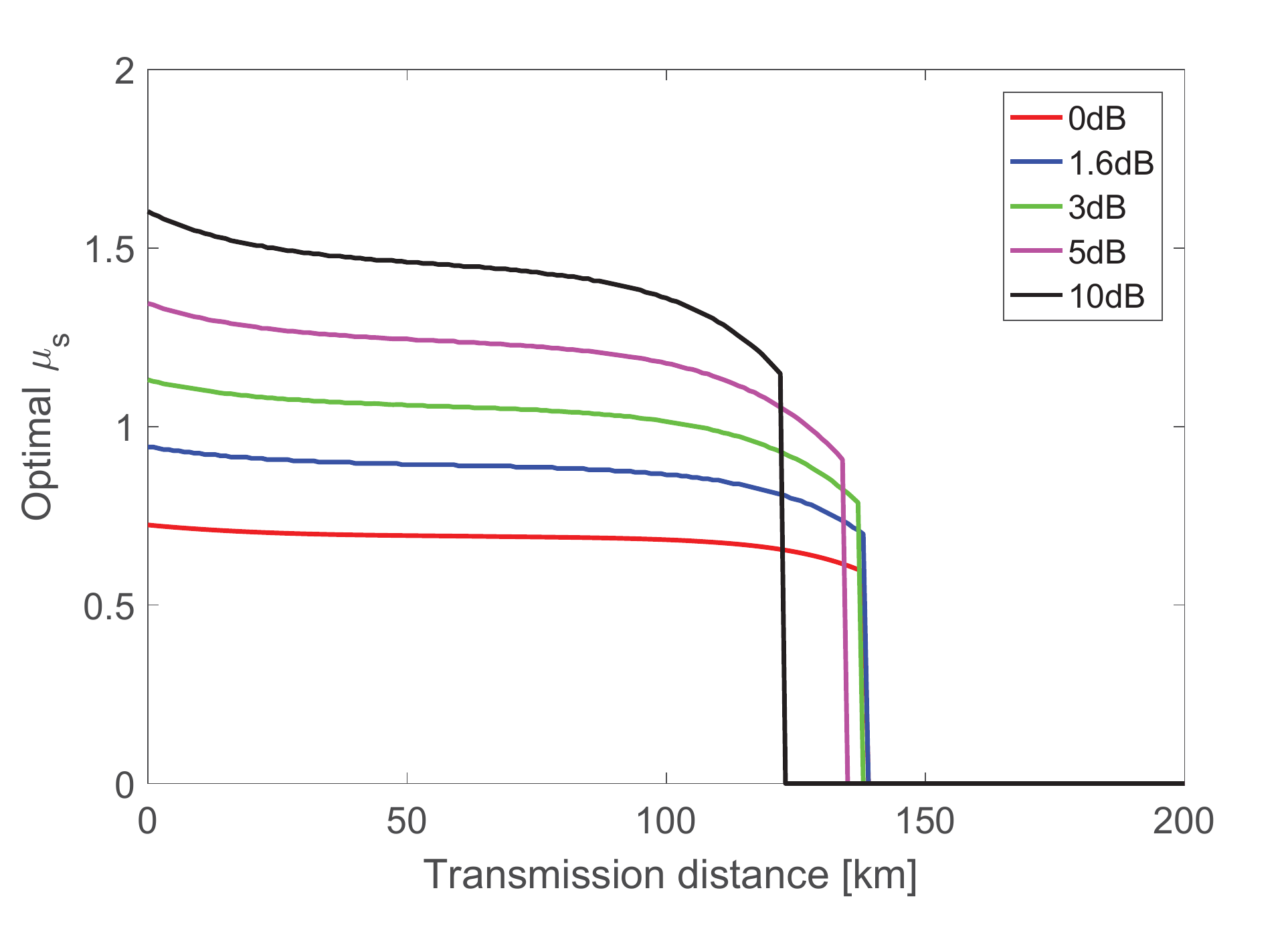}}
\subfigure[]{
\includegraphics [width=80mm,height=48mm]{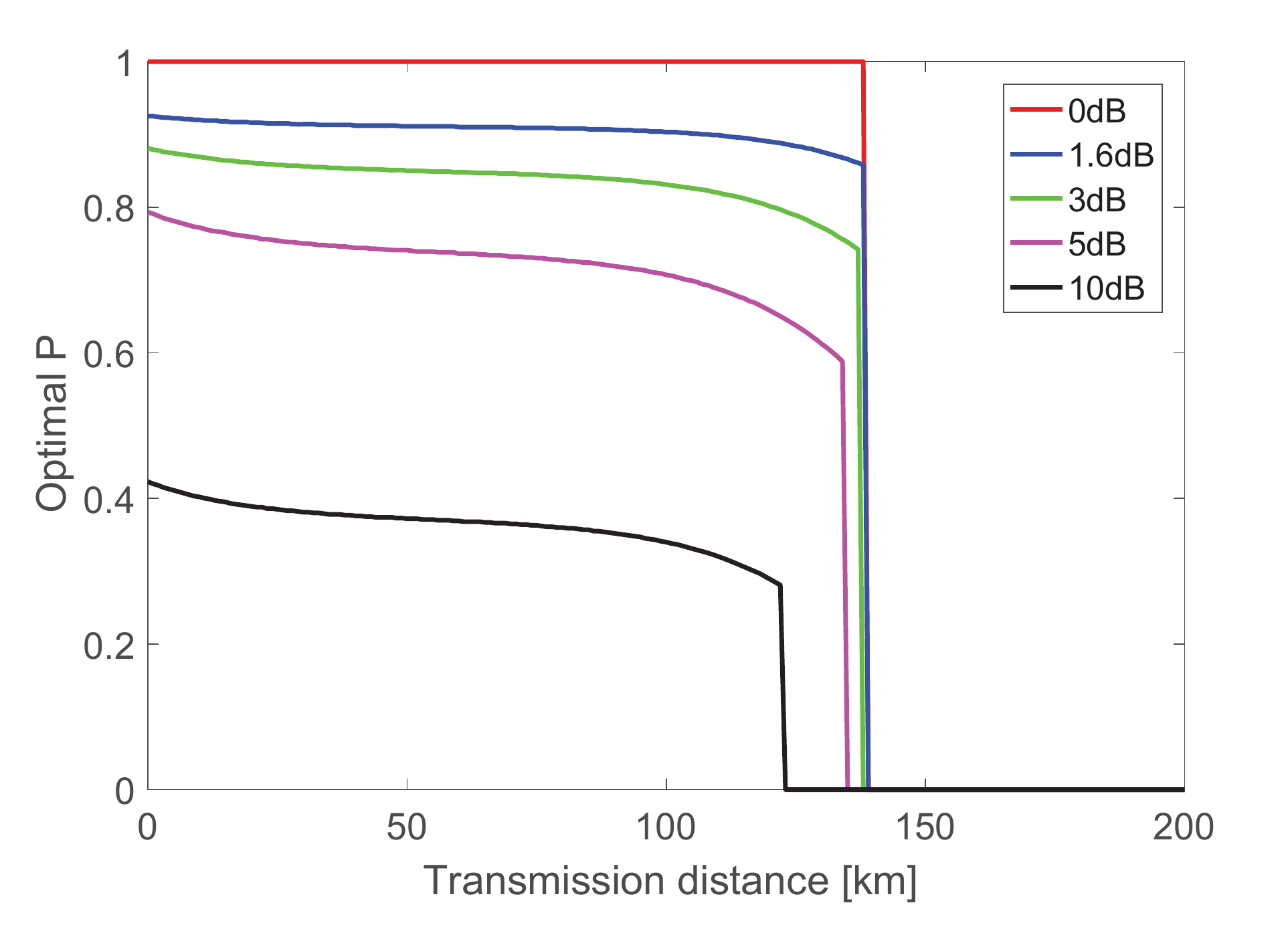}}
\subfigure[]{
\includegraphics [width=80mm,height=48mm]{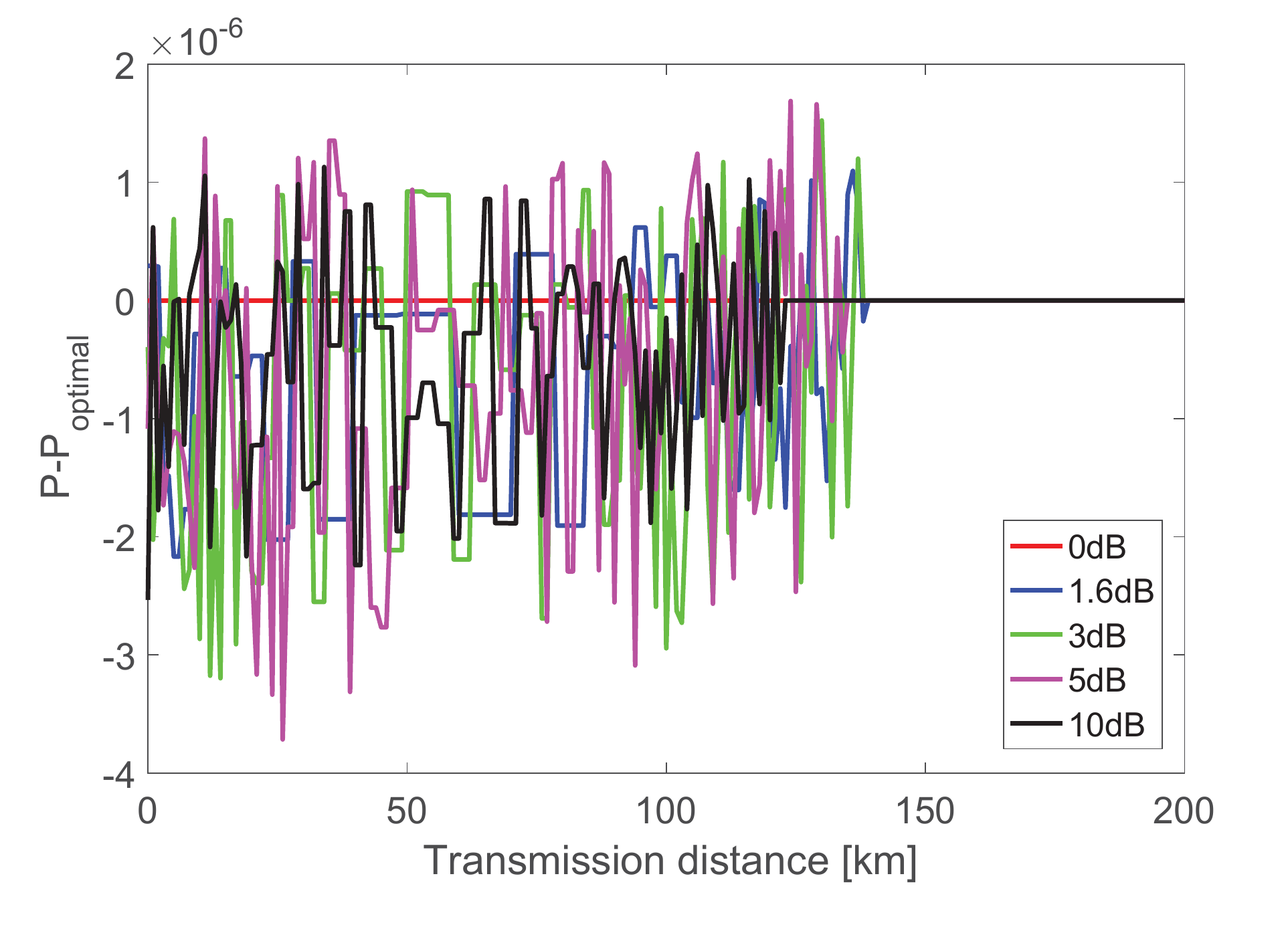}}
\caption{ Figures (a) shows the optimal values of $\mu_s$ that maximize the secret key rate according to our numerical simulations.  Figures (b) shows the optimal values of  $P$. The value of PDL increases from 0 dB to 10 dB (from top to bottom).  Figure (c) shows the deviation  between the numerically optimal value of  $P$ and the theoretical optimal value, $P_{optimal}$, obtained from Eq.(\ref{eq16}). As the deviation is rather small,  this verifies the validity of our heuristic argument. }\label{Fig3}
\end{figure}

\begin{figure}[htb]
\centering
\subfigure[$P=P_{optimal},\ \mu_s=\mu_{s,optimal} $]{\label{a}
 \includegraphics [width=80mm,height=55mm]{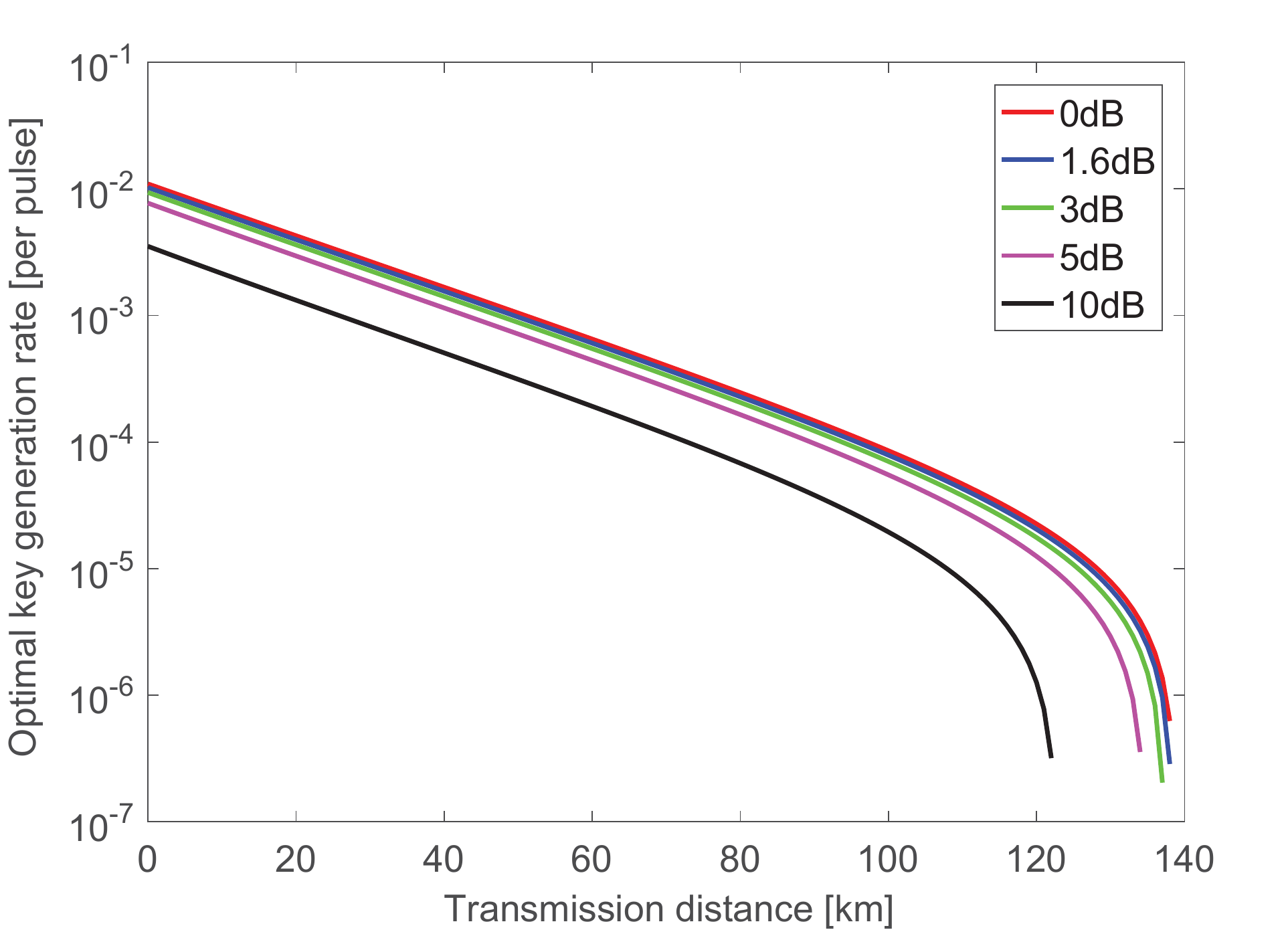}}
\subfigure[$P=1,\ \mu_s=\mu'_{s,optimal} $]{\label{b}
\includegraphics [width=80mm,height=55mm]{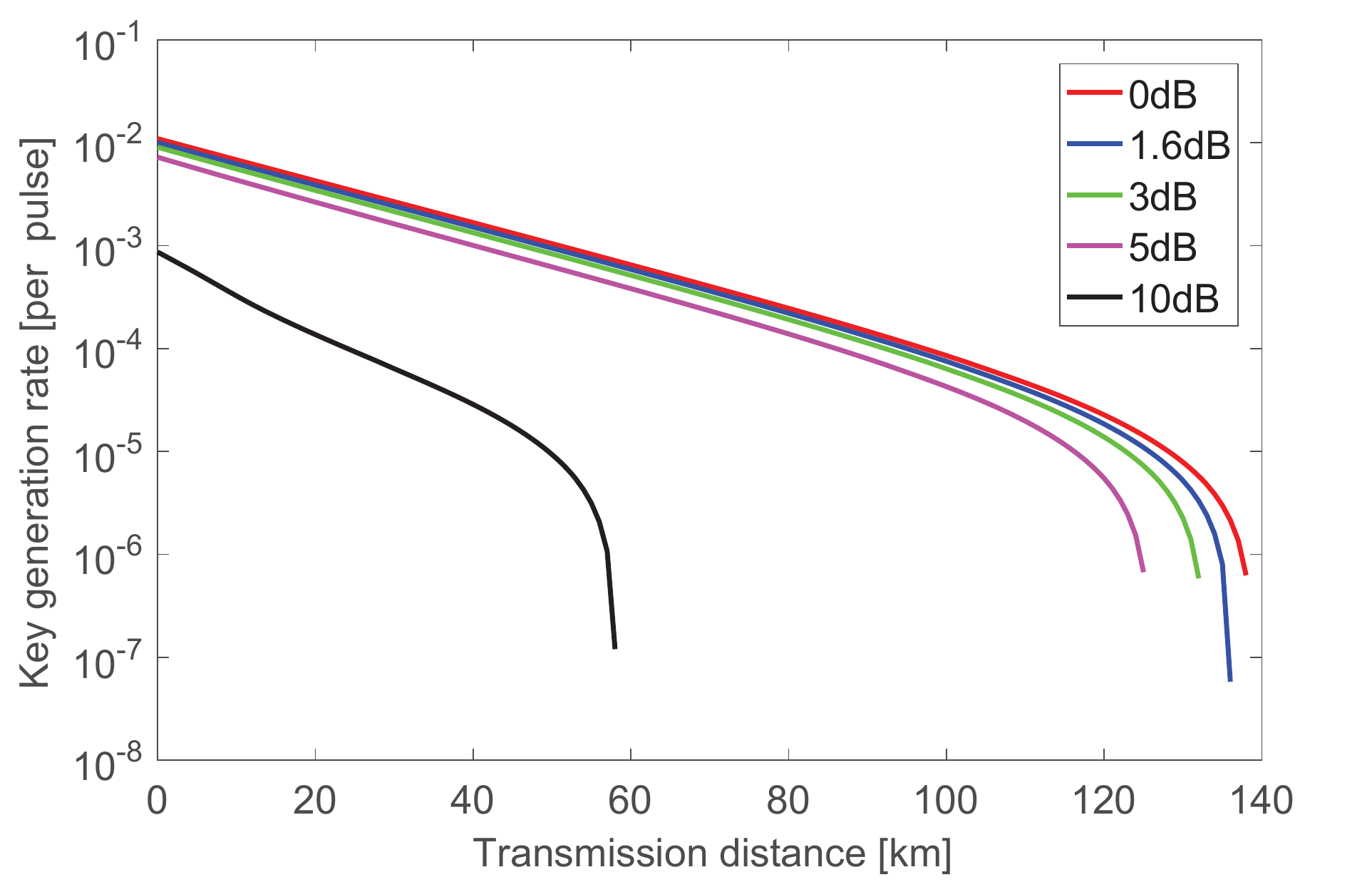}}

\caption{Comparison between (a) a QKD system  with post-selection scheme ($P=P_{optimal},\ \mu_s=\mu_{s,optimal}$)  and (b) a QKD system without post-selection scheme ($P=1, \ \mu_s=\mu'_{s,optimal}$). The secret key rate can be largely increased if the system suffers large PDL. The value of PDL increases from 0 dB to 10 dB (from top to bottom) for all figures.  }
\end{figure}\label{Fig4}

In this section, for simplicity, we  consider the asymptotic case where Alice uses an infinite number
of decoy states and an infinite number of transmission data are available for
each type of  decoy state.
\begin{table}[]
\centering
 \fontsize{10}{12}\selectfont
\caption{Experimental parameters for QKD on chip \cite{Ma2016}}
\begin{tabular}{|c|c|c|c|}

\hline
 $e_d $      & $Y_0 $  & $\eta_{Bob} $ & $\alpha(dB/km)$ \\
\hline
0.015       & $2  \times 10^{-5}$    & 0.2 &  0.2 \\
\hline

\end{tabular}
\label{table1}
\end{table}

The goal is to numerically evaluate the resulting secret key rate when one selects the optimal values for the signal intensity and the post-selection probability.  With infinite decoy states and data sizes \cite{Ma2005},  it is possible to estimate the values of $Y_1$ and $e_{1,phase}$, phase precisely. This means that for the channel model introduced above, we have that the  lower bound on the single-photon yield  is approximately $ Y_1^L=1-(1-Y_0 )(1-\eta_{sys} )$. Also, the upper bound on the single-photon phase error rate is approximately $ e_{1,phase}^U= (Y_0e_0+(Y_1-Y_0)e_d)/Y_1 $.
By substituting  the values of $Y_1^L$ and $e_{1,phase}^U$ into Eq.(\ref{eq14}) and using parameters from Table I, the secret key rate becomes a function of $P$  and $\mu_s$, that is: $R=f(P,\mu_s)$. Next, we numerically optimize the secret key $R$ over the free parameters $P$ and $\mu_s$. The results are shown in Fig.3. Figures 3 (a) and (b) show the numerical optimization results for the intensity and post-selection probability.  As mentioned before,  the optimal $\mu_s$ may exceed 1 to generate more single photons due to large PDL.  The optimal value of $P$ decreases when PDL increases, since more signals need to be discarded. Figure 3 (c) shows the deviation between  the optimal value of $P$ obtained numerically  and the theoretical value $P_{optimal}$ obtained by our heuristic argument from Eq.(\ref{eq16}). These results indicate that the optimal value of  $P$ obtained numerically  matches its theoretical value $P_{optimal}$, thus verifying the validity of our heuristic argument in Sec.\ref{Sec4}.
%%Besides, the numerical optimization of $\mu_s$ satisfies Eq.(\ref{eq18}) at the distance of 47 km.

\begin{table*}[htbp]
\centering
 \fontsize{10}{12}\selectfont
\caption{Key rate comparison at 80 km  }
\label{table3}
\begin{tabular}{|c|c|c|c|c|c|}

\hline
   PDL(dB)      & $0 $    & $1.6 $ & $3$     & $5$ & $10$ \\
\hline
  key rate with the post-selection scheme  & $10^{-3.61}$  & $10^{-3.643}$ & $10^{-3.688}$ & $10^{-3.784}$ & $10^{-4.168}$ \\
\hline
  key rate without the  post-selection scheme  &  $10^{-3.61}$  & $10^{-3.657}$ & $10^{-3.718}$ & $10^{-3.859}$ & $0$ \\
\hline
   percentage  increase &  $ 0\%$  & $3.28\%$ & $7.15\%$ & $18.9\%$ & infinity \\
\hline
\end{tabular}
\end{table*}

\begin{table*}[htbp]

  \centering
  \fontsize{10}{12}\selectfont
  \caption{Refined data analysis scheme}
    \begin{tabular}{|c|c|c|c|c|c|c|}
    \hline
    { Polarization }& \multicolumn{3}{c|}{Input}&\multicolumn{3}{c|}{ Output}\cr
  \hline
   $H$&$Q_{s,H},E_{s,H}$&$Q_{v,H},E_{v,H}$&$Q_{w,H},E_{w,H}$&$Y_{0,H}^L$& $Y_{1,H}^L$&   \cr\hline
   $V$&$Q_{s,V},E_{s,V}$&$Q_{v,V},E_{v,V}$&$Q_{w,V},E_{w,V}$&$Y_{0,V}^L$& $Y_{1,V}^L$&  \cr\hline
   $D$&$Q_{s,D},E_{s,D}$&$Q_{v,D},E_{v,D}$&$Q_{w,D},E_{w,D}$&$Y_{0,D}^L$& $Y_{1,D}^L$&$(Y_{1,D}e_{1,D})^U$  \cr\hline
   $A$&$Q_{s,A},E_{s,A}$&$Q_{v,A},E_{v,A}$&$Q_{w,A},E_{w,A}$&$Y_{0,A}^L$& $Y_{1,A}^L$&$(Y_{1,A}e_{1,A})^U$   \cr\hline

    \end{tabular}

    \label{table4}
\end{table*}

Next in Fig.4, we compare the secret key rate of a QKD system with the post-selection scheme ($P=P_{optimal}$) and that without the post-selection scheme ($P=1$). Our method without post-selection still works well in terms of small PDL. When we apply the post-selection scheme, the secret key rate can be largely increased when the source suffers large PDL.  The secret key rate comparison results are summarized in  Table \ref{table3} at the distance of 80 km.  When the value of PDL is as small as 1.6 dB, we find that the secret key rate can be improved by 3.28\% if the post-selection scheme is adopted. If the value of PDL increases, then the advantage of the post-selection scheme is more notorious. This demonstrates clearly the  benefits of using  post-selection scheme.

\section{Refined data analysis for finite decoy states with infinite data sizes}\label{Sec6}
In the previous section, our discussion was restricted to the case where Alice uses an infinite number
of decoy states. In contrast, in this section, we consider the case where the number  of decoy states is finite. For instance, in a standard two-decoy state protocol, only two decoy states are used (in addition to the signal state). The two-decoy method used to calculate the quantities, $Y_1^L,$ and $e_{1,phase}^U$,  is described in detail in \cite{Ma2005,Curty2010}. (Eg. see Eqs. (21) and (25) in \cite{Ma2005}.)

As before, we  apply a refined data analysis.
Now, for each polarization, we  apply the two-decoy method to analyze the channel. As depicted in table \ref{table4}, by separating the data from different polarizations, we now have  twelve observables: $\{Q_{n,M}, E_{n,M}\}$, where $n \in \{s,v,w\}$  denotes the intensity setting selected from the signal states ($s$) , the decoy states ($v$) and the vacuum states ($w$), and $M \in \{H,V,D,A\}$. Now we are able to estimate a lower bound on the single-photon yields for different polarizations based on the specific intensities. Moreover, note that for this it is not necessary to use the assumptions given by Eqs. (3)-(4)-(7)-(10) but our approach is general.

\begin{figure}[htp]
\centering
 \includegraphics[width=80mm,height=55mm]{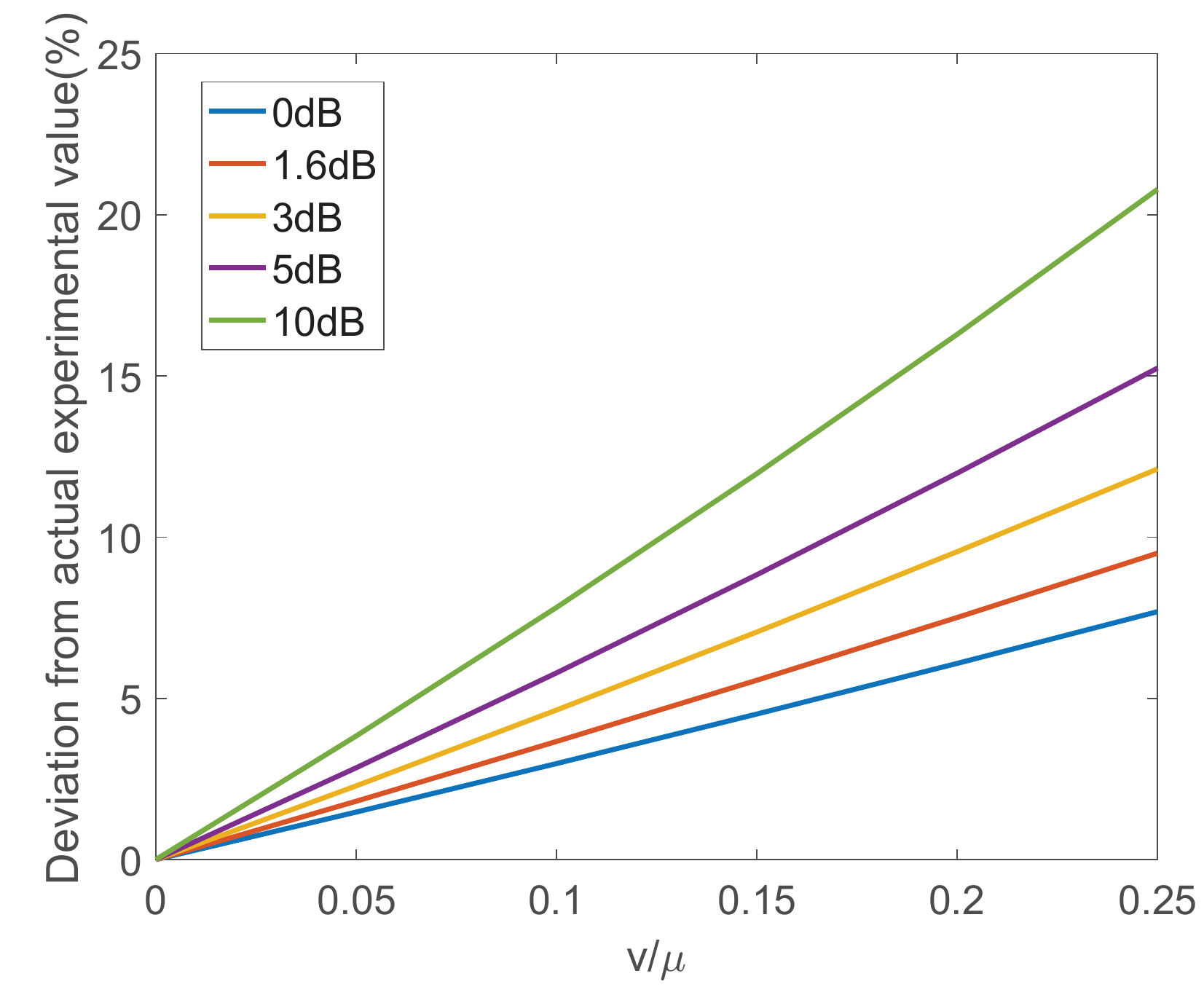}

\caption{ The deviation of $Y_1$ is proportional to the intensity of the  decoy states. Large PDL values lead to large deviations. The value of PDL increases from 0 dB to 10 dB (from bottom to top).}\label{Fig5}
\end{figure}

\begin{figure}[!htp]
\centering
\subfigure[]{\label{a}
 \includegraphics [width=80mm,height=55mm]{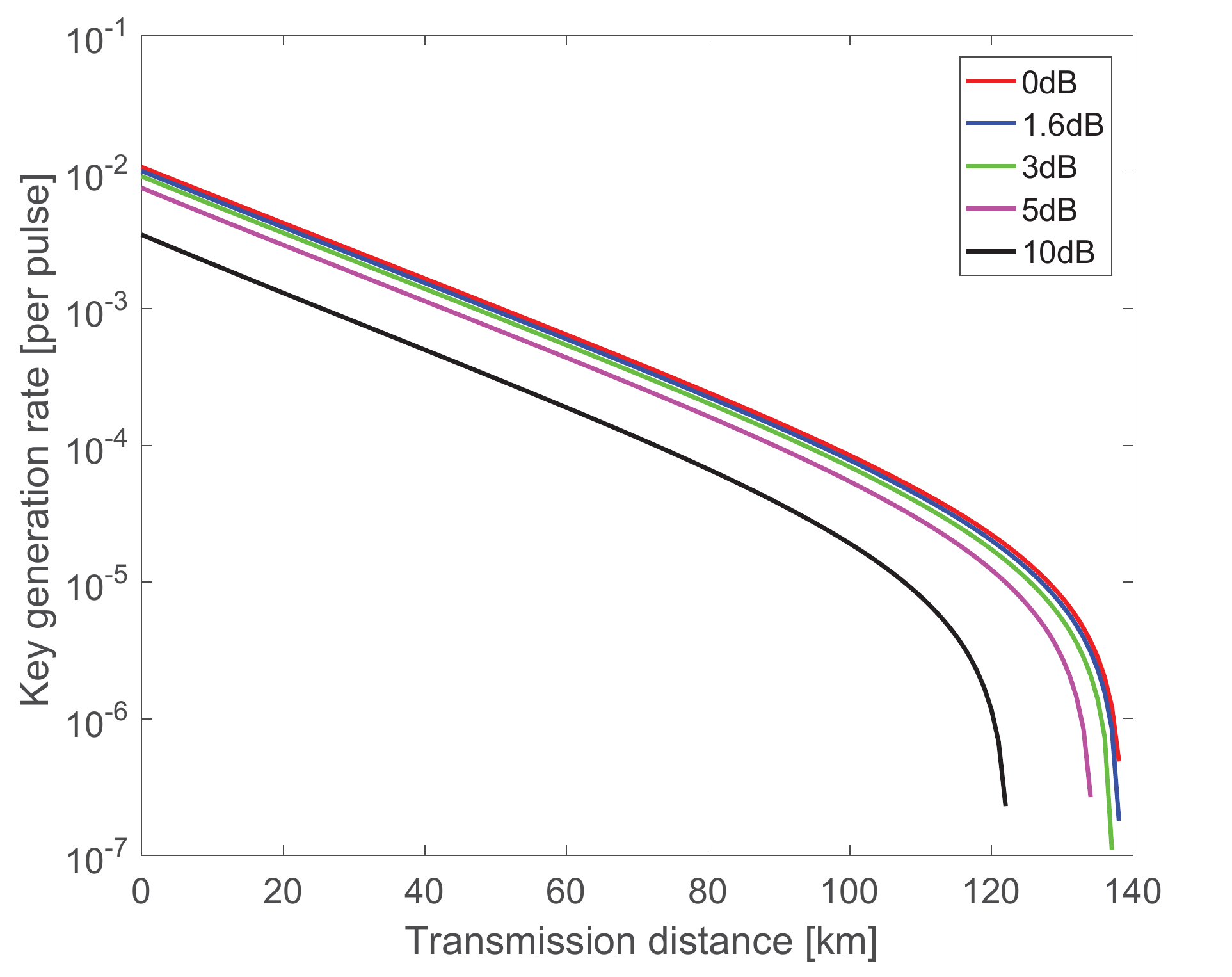}}
\subfigure[]{\label{b}
\includegraphics [width=80mm,height=55mm]{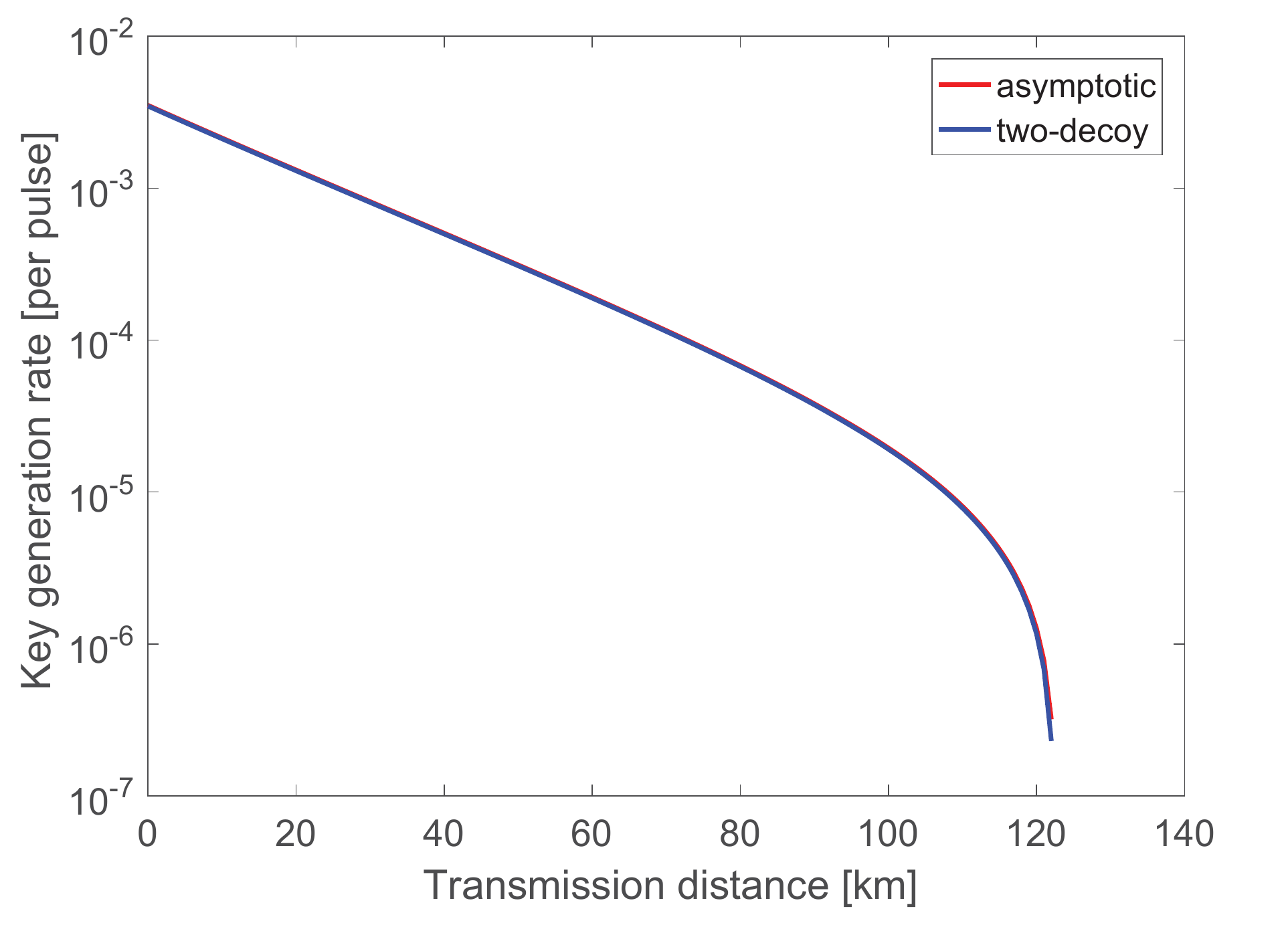}}

\caption{ Here, we consider the two-decoy state method (decoy+vaccum) with $\mu_w=0$ and numerically optimize the secret key rate over the free parameters, $\mu_s$ and $\mu_v$. Also, we use the optimal value of $P$ given by Eq. (18). (a) We estimate the secret key rate with different PDL values.  The value of PDL increases from 0 dB to 10 dB (from top to bottom). (b) We compare the two-decoy state method (bottom) with the asymptotic case (top) described in Fig.4 (a). The value of PDL is 10 dB. The two-decoy state method works well, since the resulting secret key rate is close to the asymptotic case.  }\label{Fig6}
\end{figure}
Based on the secret key rate formula given by Eq.(\ref{eq14}), we need to estimate a lower bound on $Y_1$ and an upper bound on $e_{1,phase}$.
By using the refined data analysis, we have:
\begin{gather}
  Y_1^L=\frac{Y_{1,H}^L+Y_{1,V}^L}{2}, \\ \notag
  e_{1,phase}^U=\frac{(Y_{1,D}e_{1,D})^U+(Y_{1,A}e_{1,A})^U}{Y_{1,D}^L+Y_{1,A}^L}.  \label{eqsp}
\end{gather}

 First, we  show the estimated values of  $Y_1^L$  when the source suffers different PDLs. For illustrative purposes, we consider $Y_1=Y_0+\eta_{sys}$  to be the actual experimental value(i.e., we consider that the term $Y_0*\eta_{sys}$ is neglectable).  Therefore, the relative deviation is given by:
\begin{gather}
  \beta_{Y_1} = \frac {Y_1-Y_1^L}{Y_1}.  \label{eq19}
\end{gather}
Moreover, we also use the experimental parameters for the QKD chip shown in Table \ref{table1} and change the intensity of the weak decoy state to see how it affects our estimation. As shown in Fig.\ref{Fig5}, the deviation of $Y_1$ is proportional to the intensity of the weak decoy state in the case of infinite data sizes and the trend of the deviation follows the PDL. That is, when  the PDL at the source increases,  the deviations becomes larger. This is expected since a larger deviation corresponds to a smaller secret key rate.

Next, we consider the secret key rate. For this, we numerically optimize the secret key rate over the free parameters \{$\mu_s$, $\mu_v$\}. By computing $Y_1^L$ and $e_{1,phase}^U$ and substituting their values in Eq.(\ref{eq14}), we obtain the final result of the secret key rate. As depicted in Fig.6, the two-decoy state method works well compared to the asymptotic case described in Sec.V.

\section{statistical fluctuations for finite data sizes}\label{Sec7}
In this section, we  first define the security criteria that we use \cite{Renner2005,Lim2014}. For some small errors, $\varepsilon_{\rm cor},\ \varepsilon_{\rm sec} > 0$, we say that a QKD protocol is $\varepsilon_{\rm cor}\ + \ \varepsilon_{\rm sec}$ secure if it is $\varepsilon_{\rm cor}$ correct and $\varepsilon_{\rm sec}$ secret. The former is satisfied if  Alice's and Bob's secret keys are identical except with a small probability $\varepsilon_{\rm cor}$. The latter is satisfied if $||\rho_{AE}\ -\ U_A \ \otimes \ \rho_E||_1/2 \ \leq \ \varepsilon_{\rm sec}$, where $\rho_{AE}$  is the classical-quantum state describing the joint state of $S_A$ and $E$, and $U_A$ is the uniform mixture of all possible values of $S_A$, where $S_A$ is the secret key of Alice and $E$ denotes Eve's quantum system. Importantly, this secrecy criterion guarantees that the protocol is universally composable, i.e., the pair of secret keys can be safely used in any
cryptographic task. Conditioned on passing the tests in the error-estimation and error-verification steps, the length of the $\varepsilon_{\rm cor}\ + \ \varepsilon_{\rm sec}$ secure key  in the Z basis is given by \cite{Renner2005,Lim2014}:
 \begin{gather}
  l\ \geq \  s_{z,s,1}^L [1- h(e_{x,1}^U)] -\ \lambda_{\rm EC}\\           \notag
   -\ 6\ {\rm log}_2 \frac{21}{\varepsilon_{\rm sec}}\ -\  {\rm log}_2 \frac{2}{\varepsilon_{\rm cor}},
\end{gather}
where $s_{z,s,1}^L$ is a lower bound on the number of single-photon events for signal states in the Z basis that contribute to the sifted key, and  $e_{x,1}^U$ is an upper bound on the single-photon phase error rate estimated from the X basis events. $\lambda_{\rm EC}$ is the syndrome information declared  for error correction.
With $l$, the secret key rate is given by $R^l\ =\ l/N$ with $N$ denoting the total number of signals (optical pulses) sent by Alice.

For a  QKD system with a two-decoy state method (decoy+vaccum),
let the symbols $N_s,\ N_v$and $\ N_w$ denote  the number of pulses sent by Alice
in the three intensities. Then the total number of pulses sent by Alice is given by:
\begin{equation}\label{eq20}
  N=N_s+N_v+N_w.
\end{equation}
The probability to choose the intensity setting $n \in \{s,v,w\}$ is then given by $P_n=N_n/N$.
Alice and Bob both use the Z basis for key generation and the X basis to estimate the phase error rate. The basis choice  probability is set for simplicity to be $P_X=P_Z=1/2$. The polarization choice  probability is $P_M=1/2$ for each basis, where $M \in \{H,V,D,A\}$. The number of pulses for different polarization, $N_{n,M}$, is given by: $N_{n,M}=N_n/4$. Here, we recall that we apply the post-selection scheme for the signal state in the Z basis, i.e., vertical polarization. Thus, we can pre-choose a random data size, $N_{s,V}=P\times N_s/4$, to keep this data while we discard other data in the vertical polarization. In fact, the signal state intensity is typically used much more frequently than the two decoy state intensities, so the statistical fluctuations are still small after considering the post-selection data size effect. Then the number of single-photon events from signal states has the form: $ s_{z,s,1}^L=\ Q_{1,s,H}^LN_{s,H}+\ Q_{1,s,V}^LN_{s,V}$, where $Q_{1,s,H}^LN_{s,H}$ and $Q_{1,s,V}^LN_{s,V}$ are, respectively, the  lower bounds on the number of  single-photon events in horizontal and vertical polarizations which are detected by Bob.

Based on the table \ref{table4}, we have 12 experimental observables: $Q_{n,M}N_{n,M}$ and $\ Q_{n,M}E_{n,M}N_{n,M}$. For a finite data size, due to Hoeffding's inequality \cite{Lim2014,Hoeffding1963}, the experimental values $Q_{n,M}N_{n,M},\ Q_{n,M}E_{n,M}N_{n,M} $ and the expected values $\bar{Q}_{n,M}N_{n,M},\ \bar{Q}_{n,M}\bar{E}_{n,M}N_{n,M} $ satisfy:
\begin{gather}
   |Q_{n,M}N_{n,M}-\bar{Q}_{n,M}N_{n,M}|\leq \Delta(Q_MN_M, \varepsilon_{\rm sec}),\notag \\ \label{eq22}
  |Q_{n,M}E_{n,M}N_{n,M}-\bar{Q}_{n,M}\bar{E}_{n,M}N_{n,M}|\leq \Delta(Q_ME_MN_M,\varepsilon_{\rm  sec}),
\end{gather}
with probability at least $1 - 2 \varepsilon_{\rm sec} $ , where $Q_MN_M$ and $Q_ME_MN_M$, respectively,  denote all detection events and errors in the $M$ polarization, with $M \in \{H,V,D,A\}$, and $\Delta(x,\ \varepsilon):=\sqrt{x/2{\rm ln}(1/\varepsilon)}$. Note that, compared to the standard error analysis presented in \cite{Ma2005}, the analysis introduced in \cite{Lim2014} considers general attacks.

For the evaluation, we  use the chip-based QKD parameters shown in Table \ref{table1} and fix $\varepsilon_{\rm cor}\ = \ \varepsilon_{\rm sec} \ =\ 10^{-10} $. We pick  the total data size to be between $N=10^{10}$ and $10^{14}$.
We numerically optimize the secret key rate $R^l$ over the free parameters $\{\mu_s,\ \mu_v ,\ P_s,\ P_v\}$.
 The optimization result is depicted in Fig.7. In terms of the secret key rate, it is advantageous to select a larger data size, since a smaller data size corresponds to larger statistical fluctuations. Besides, the secret key rate is approaching the asymptotic case when  the data size increases. Even in the presence of a PDL as high as 10 dB and a data size of $10^{10}$, the maximum distance  is about  100 km. This suggests the effectiveness of our post-selection scheme.

 \begin{figure}[htbp]
\centering
 \includegraphics[width=80mm,height=55mm]{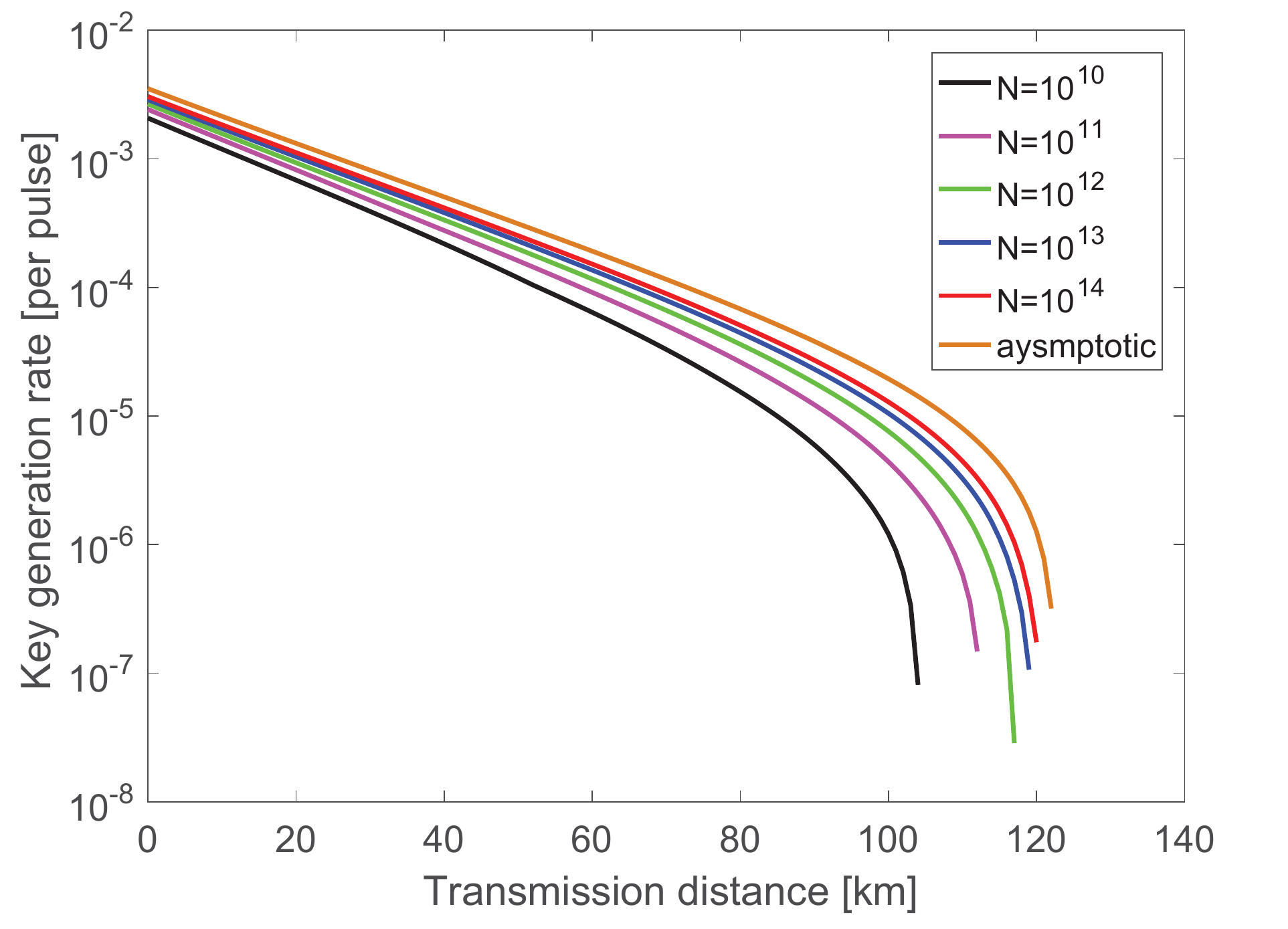}

\caption{Simulation result of the finite data size effect. We select  the total number of pulses to be between $N=10^{10}$ and $10^{14}$ (from left to right). The value of PDL is 10 dB.  As expected, the secret key rate is approaching the asymptotic case if we increase the data size.  }\label{Fig7}
\end{figure}

\section{conclusion}
\label{Sec8}
We have studied the security of  QKD in the presence of PDL. The secure key rate of QKD system will be reduced due to a polarization dependence loss in the source. Because of the overestimation of the key rate, Alice and Bob will generate a key that is not guaranteed to provide information-theoretic security.  Here, we first restore the security by generating the key from the single-photon components whose density matrices are maximamlly mixed. In \cite{Gottesman2004}, the single photons  which are basis independent are regarded to be untagged and secure qubits. Our security proof follows this idea and regards the single photons without polarization dependence in the source as untagged and secure qubits. Next, when there is a large PDL in the system, the maximum transmission distance becomes rather short, as the imbalance of the single photon portion may leak more information to Eve. Trying to balance the single photon portion in the encoding basis, we have proposed a post-selection scheme that discards the signals in a polarization with a smaller PDL or without PDL. Given the post-selection scheme and  refined data analysis, we have  numerically optimized the intensity settings and the post-selection probability. Finally, we have studied the decoy state method and the finite data size effect. The two-decoy state method works well in our scheme. By increasing the total data size, one can achieve the asymptotic secret key rate. In summary, our work provides a simple software solution that compensates for PDL in silicon photonics QKD, thus paving the way to low-cost high-speed QKD transmitters based on silicon photonics.

\section{acknowledgements}\label{Sec9}
We acknowledge enlightening discussions with many colleagues including Joyce Poon, Gligor Djogo, Yisu Yang and Wengyuan Wang.
We also acknowledge the financial support the Natural Sciences and Engineering Research Council of Canada (NSERC), the National Research Council (NRC) of Canada and Huawei Technologies Canada Co., Ltd.  M. Curty acknowledges support from the Spanish Ministry of Economy and Competitiveness (MINECO), the Fondo Europeo de Desarrollo Regional (FEDER) through  grants TEC2014-54898-R and TEC2017-88243-R, and the European Union's Horizon 2020 research and innovation programme under the Marie Sk\l{}odowska-Curie grant agreement No 675662 (project QCALL). F. Xu acknowledges support from 1000 Young Talent Program of China, the Fundamental Research Funds for the Central Universities of China.

\end{document}